\documentclass[12pt]{article}
\usepackage{amsfonts}
\newsavebox{\uuunit}
\sbox{\uuunit}
    {\setlength{\unitlength}{0.825em}
     \begin{picture}(0.6,0.7)
        \thinlines
        \put(0,0){\line(1,0){0.5}}
        \put(0.15,0){\line(0,1){0.7}}
        \put(0.35,0){\line(0,1){0.8}}
       \multiput(0.3,0.8)(-0.04,-0.02){12}{\rule{0.5pt}{0.5pt}}
     \end {picture}}

\begin{document}
\begin{flushright}
FTUV-07-0109 \hskip .4cm IFIC/07-09
\\ Jan. 9 -- Feb. 13, 2007
\end{flushright}
\vspace{.5cm}

\begin{center}
 {\bf \large BPS preons in M-theory and supergravity}\footnote{
 Talk delivered at the Workshop of the RTN network {\it Constituents, fundamental forces
 and Symmetries of the Universe}, Napoli October 9-13, 2006, to
 appear in the proceedings (Fortschritte der Physik).}

\bigskip

{ Igor A. Bandos$^{\dagger,\,\ddagger}$, Jos\'e A. de
Azc\'arraga$^\dagger$}

\bigskip

$^\dagger$ {\it Department of Theoretical Physics, Valencia
University, and IFIC (CSIC-UVEG), 46100-Burjassot (Valencia),
Spain}

$^\ddagger${\small\it Institute for Theoretical Physics, NSC KIPT,
 61108 Kharkov, Ukraine}

\end{center}

\bigskip

\begin{center}
{\bf Abstract}
\end{center}
\begin{quotation}
After introducing the notion of BPS preons as the basic
constituents of M-theory, we discuss the recent negative results
in the search for solutions of the $D$=10 and $D$=11 supergravity
equations preserving 31/32 supersymmetries {\it i.e.}, of preonic
solutions. The absence of these supergravity preonic solutions may
point out to a pure quantum nature of BPS preons, manifesting
itself in the need of incorporating quantum (stringy/M-theoretic)
corrections to the supergravity equations.
\end{quotation}

\bigskip

\section{Introduction}
{\it BPS preons} are BPS states preserving all but one
supersymmetries. They were introduced six years ago \cite{BPS01},
when the list of known supersymmetric solutions of $D$=11 and
$D$=10 type II supergravities only contained solutions preserving
$16$ or less supersymmetries, or all the 32. These solutions
described $k/32$ BPS states with $k\leq 16$ plus the fully
($32/32$)-supersymmetric vacua, the set of which contained then
the Minkowski and some $AdS_d\otimes S^{D-d}$ spaces \cite{SGsol}
as well as the $pp$-wave backgrounds \cite{JK-G83-Hull84}.

Although it was already clear that BPS states with $k>16$ were not
forbidden by the M-theory algebra \cite{BL98, G+H99}, no
supersymmetric solutions with $16<k< 32$ were known in 2001. As a
result, the existence of $(31/32)$-supersymmetric or {\it preonic}
solutions of the supergravity equations was hardly expected at the
time, and the possibility of a {\it preon conspiracy} by which
only composites of some number of preons coud be `observed' as
supergravity solutions, but not the single preons themselves, was
already discussed in \cite{BPS01,BPS03}. The expectancy of finding
preonic, $31/32$ supersymmetric solutions increased when some
examples preserving more than one-half, {\it i.e.} having more
than 16 supersymmetries were found, mainly as plane wave solutions
(see \cite{ppW} and the references in \cite{Duff02, BPS03}).
Nevertheless, the search for 31/32 supersymmetric solutions of the
`free' (no sources) supergravity equations, which we review here,
has only produced negative results. Indeed, for type II
supergravities \cite{GGPR-IIB-06, BdAV06} and for simply connected
solutions of $D$=11 supergravity \cite{GGPR-11-06}, the existence
of preonic solutions has now been ruled out. Nevertheless, these
`no go' conclusions may change if quantum M-theoretic or `stringy'
corrections are taken into account. Moreover, the preon conspiracy
does not preclude the preon conjecture \cite{BPS01}, since preons
were introduced as M-theoretical objects and fundamental
constituents of M-theory as a whole, rather than as specific
solutions of supergravity, its low energy limit.

Let us begin by reviewing the definition of M-theoretic BPS preons
\cite{BPS01,30/32,BPS03}.

\section{ BPS preons as M-theory constituents}

A BPS preon \cite{BPS01} is an M-theory BPS state preserving all
but one supersymmetries, namely $31$ out of $32$. It is
equivalently labelled as $|BPS\; preon>:=|31/32\; susy>:= |31/32\;
BPS>$ (we note, however, that the preon concept is not restricted
\cite{BPS01} to $D$=11). Denoting the $32$ component supersymmetry
generator by $Q_\alpha$, the fact that a BPS state preserves $k$
supersymmetries is expressed as
\begin{eqnarray}
\label{Preons}
\fbox{$\; \epsilon_I^{\;\alpha} Q_{\alpha}|{\it k\over 32} \;
BPS>= 0\; $}\; , \quad {}^{\alpha= 1,\ldots , 32\;}_{
  I=1, \ldots , k \;}\quad  (a)\; ; \quad &
 \fbox{$\;  |BPS\; preon>= |{\it 31\over 32} \; BPS> \;$}\quad (b)\;, \qquad
\end{eqnarray}
where $\epsilon_I^{\;\alpha}$ are the $k$ bosonic spinors ($k=31$
for a BPS preon) that characterize the $k$ preserved
supersymmetries. They will correspond to Killing spinors in the
supergravity solutions case below.

The only supersymmetry broken by a BPS preon can be characterized
by one bosonic spinor. This can be, for instance, a bosonic spinor
$u^\alpha$ such that $u^\alpha Q_\alpha |BPS\; preon>\neq 0$.
Together with eq. (\ref{Preons}), this implies that $det
\{u^\alpha\, , \, \epsilon_I^{\;\alpha}\}\not= 0$, which expresses
that $u^\alpha\,$ is linearly independent of the $31$ spinors
$\epsilon_I^{\;\alpha}$. However, it is more convenient to
characterize a BPS preon state by another spinor $\lambda_\alpha$,
which is orthogonal to all the  bosonic spinors
$\epsilon_I^{\;\alpha}$ that determine the 31 supersymmetries
preserved by the BPS preon,
\begin{eqnarray}\label{epXl=0}
 \fbox{$\epsilon_I^{\;\alpha} \lambda_\alpha= 0\;$}\;   , \qquad  {}^{\alpha= 1,\ldots , 32\;}_{
  I=1, \ldots , 31 \;} \quad
 \qquad  \Rightarrow  \; & |BPS\; preon>= |{\it 31\over 32} \; BPS>= |\lambda > \; .
\end{eqnarray}
Thus, the non-zero result of the action $Q_{\alpha}|BPS\; preon>$
of $Q_{\alpha}$ on a preon state (\ref{Preons}), has to be given
in terms of the bosonic spinor $\lambda_\alpha$ that characterizes
the preon. To express this explicitly, one has to introduce
another state $|BPS\; preon^f>= |{\it 31\over 32}^f \; BPS>=
|\lambda\; {}^f
> $ with opposite Grassmann parity, fermionic {\it if}
the 31/32 state is considered to be bosonic as it is usually (but
not necessarily) the case. Then, the pair $\{\,
|\lambda\;>\,,\,|\lambda\; {}^f
>\,\}$ determines an ultrashort {\it BPS preonic multiplet}
\cite{BdA06-J}, on which the action of the supersymmetry generators
reads
\begin{eqnarray}\label{PreonsUsusy}
 Q_{\alpha}|\lambda\;>\;=\; \lambda_\alpha |\lambda\; {}^f > \;  \quad (a)
 \quad , \qquad
Q_{\alpha}|\lambda\; {}^f >\;=\; \lambda_\alpha |\lambda\; >\; \quad
(b)\quad.
\end{eqnarray}
Notice that the bosonic and fermionic preonic states enter
(\ref{PreonsUsusy}) in a completely symmetric manner, so that one
might also think of identifying the BPS preon with a fermionic
state \cite{BdA06-J}. This requires further study, and in this
paper we shall concentrate on bosonic BPS preons as they might be
described by purely bosonic $31/32$ supersymmetric solutions of
the supergravity equations and their M-theoretic generalizations
{\it i.e.}, those incorporating quantum `stringy'
($(\alpha^\prime)^3$) corrections.

Applying the supersymmetry generator to eq. (\ref{PreonsUsusy}a) and
using (\ref{PreonsUsusy}b), one finds that
 $Q_\beta Q_\alpha|\;\lambda \,>\;=
 \; \lambda_\beta\lambda_\alpha |\; \lambda \, > $,
 which leads to another equivalent definition of a BPS
 preon state \cite{BPS01,30/32,BdA06-J},
\begin{eqnarray}
\label{def2-Preon} \fbox{$P_{\alpha\beta}|BPS\; preon>\;=\;
\lambda_\alpha\lambda_\beta \, |BPS\; preon> \;$}\;  . \qquad
\end{eqnarray}
The symmetric generalized momentum $P_{\alpha\beta}$ operator above
is the bosonic generator in the superalgebra
 \begin{eqnarray}
 \label{M-alg}
 {} \{Q_\alpha \, , \, Q_\beta \} = 2P_{\alpha\beta}\; , \qquad [
P_{\alpha\beta} \, , \, P_{\gamma \delta}] = 0\; , \qquad [
P_{\alpha\beta} \, , \, Q_{\gamma}] = 0 \; , \qquad \alpha
  =1, \ldots , 32\quad ,
 \end{eqnarray}
which has a central extension structure and is associated to the
maximally extended rigid tensorial superspace
$\Sigma^{(n(n+1)/2|n)}$, where $n$ is the dimension of the
appropriate spinor (32 for $D$=11).

The second definition (\ref{def2-Preon}) of a BPS preon indicates
that any BPS state $|k/32\; BPS>$ preserving $k$ supersymmetries
may be considered \cite{BPS01} as  a composite of $\tilde{n}=32-k$
BPS preons, $\;|BPS\; preon \; ^{l}>= |\lambda \; ^{l}>\;$,
$\;l=1,\ldots, \tilde{n}\;$ :
\begin{eqnarray}\label{composed}
& |{k\over 32} \;  BPS>   = | \lambda^{(1)}> \otimes \ldots
\otimes |\lambda^{(32-k)}> \equiv \bigoplus_{r=1}^{\tilde{n}}|
\lambda^{(r)}>\; , \qquad \tilde{n}=32-k \; .
 \end{eqnarray}
To conclude this, we begin by noticing that the eigenvalue matrix
$p^{k}_{\alpha\beta}$ of $P_{\alpha\beta}$ for a BPS state
preserving $k$ supersymmetries $|{k\over 32} \; BPS>$ has rank
$\tilde{n}=32-k$ (one for a preon).  This is expressed by the
equation
\begin{eqnarray}
\label{pk=sum} &
p^{k}_{\alpha\beta}= \lambda_\alpha^1\lambda_\beta^1 + \ldots +
\lambda_\alpha^{\tilde{n}}\lambda_\beta^{\tilde{n}} =
\sum_{r=1}^{\tilde{n}=32-k}\lambda_\alpha^{r}\lambda_\beta^{r}\; ,
\qquad P_{\alpha\beta}|{k\over 32} \;  BPS>=
p^{k}_{\alpha\beta}|{k\over 32} \; BPS>\; , \quad
\end{eqnarray}
where $\lambda_\alpha^{r}$ are the $\tilde{n}$ bosonic spinors that
characterize the supersymmetries broken by the ${k/32}$ state and
that appear in the expression
\begin{eqnarray}\label{32-k-br}
& Q_\alpha |{k\over 32} \;  BPS>   =\lambda_\alpha^{1}  |f^1> +
\ldots + \lambda_\alpha^{\tilde{n}}  |f^{\tilde{n}}>  =
\bigoplus_{r=1}^{\tilde{n}} \lambda_\alpha^{r} |f^r> \; , \qquad
\tilde{n}=32-k\; ,
\end{eqnarray}
where $|f^1>$, ..., $|f^{\tilde{n}}>$ are some fermionic states.
The additive structure of the generalized momentum eigenvalue
matrix $p^{k}_{\alpha\beta}$ of a $|{k\over 32} \; BPS>$ state,
eq. (\ref{pk=sum}), then shows that such a BPS state may be
treated as a composite \cite{BPS01} of a $\tilde{n}=32-k$
independent BPS preons, each of them characterized by an
independent spinor $\lambda_\alpha$, eq. (\ref{composed}). Thus,
the number ${\tilde n}$ of {\it broken} supersymmetries ${\tilde
n}=32-k$ is the number of preons that compose the BPS state; each
preon breaks one supersymmetry (see \cite{30/32,BPS03} for further
discussion), and a fully supersymmetric vacuum contains no preons.

The $\tilde{n}= 32-k$ {\it preonic spinors}  $\lambda_\alpha^{r}$
that characterize the preons making the $|{k\over 32} \; BPS>$
state are orthogonal to the $k$ (Killing) spinors
$\epsilon_{\!_{I}}{}^\alpha$ that characterize the $k$
supersymmetries preserved by it, eq. (\ref{Preons}a),
\begin{eqnarray}\label{epkXl=0}
 \fbox{$\; \epsilon_{\!_{I}}^{\;\alpha} \lambda_\alpha{}^r= 0\;$}\;   , \qquad
 \alpha= 1,\ldots ,
 32\; , \qquad
  I=1, \ldots , k \; , \quad r= 1,\ldots \,  ,
  \tilde{n}\quad (\tilde{n}=32-k)  \;.
  \quad
\end{eqnarray}
To conclude this section we note that the fermionic states $|f>$
in (\ref{32-k-br}) differ from the original bosonic
k/32-supersymmetric BPS state $|{k\over 32} \; BPS>$
(\ref{composed}) by replacing one of the bosonic preons
$|\lambda>$ in (\ref{composed}) by its fermionic superpartner
$|\lambda{}^{f}>$, namely
 $\;|f^r>:= |\lambda^{(1)}> \otimes|\lambda^{(2)}> \otimes
 \ldots \otimes|\lambda^{(r)}{}^f>\otimes\dots \otimes |\lambda^{(32-k)}>$.

\section{The moving G-frame method and generalized holonomy  in supergravity}

In supergravity, the $k$ bosonic spinors
$\epsilon_{\!_{I}}^{\;\alpha}$ of the supersymmetries preserved by
a $k\over 32$ BPS solution (eq. (\ref{Preons}a)) are Killing
spinors obeying the equation,
\begin{eqnarray}\label{Killing} \fbox{$\; {\cal D} {\epsilon}_{I}{}^\alpha \; := D
{\epsilon}_{I}{}^\alpha - {\epsilon}_{I}{}^\beta
{t}{}_\beta{}^\alpha\, = 0\,$} \, , \quad &
D{\epsilon}_{I}{}^\alpha := d {\epsilon}_{I}{}^\alpha -
{\epsilon}_{I}{}^\beta \omega_\beta{}^\alpha\, ,\quad
\omega_\beta{}^\alpha:= {1\over 4}
\omega^{ab}\Gamma_{ab}{}_\beta{}^\alpha\, , \quad
\end{eqnarray}
where $D=e^aD_a$ is the standard (Lorentz) covariant derivative,
$\omega^{ab}=-\omega^{ba}=e^c\omega_c{}^{ab}$ is the one-form spin
connection and $t{}_\beta{}^\alpha\,:= e^a t_a{}_\beta{}^\alpha\,$
is the tensorial contribution to the {\it generalized} connection
$w{}_\beta{}^\alpha= \omega{}_\beta{}^\alpha + t{}_\beta{}^\alpha
$ defining the {\it generalized} covariant derivative ${\cal
D}=d-w:= D-t$. This tensorial contribution  is constructed from
the gauge field strength(s) and, in $D$=10, also includes the
derivatives of the axion and dilaton. For instance, in the case of
$D=11$ supergravity the tensorial contribution reads
\begin{eqnarray}\label{t11D}
 & t_\beta{}^\alpha = {i\over 18} e^a
\left(F_{a[3]} \Gamma^{[3]} + {1\over 8} F^{[4]}
\Gamma_{a[4]}\right){}_\beta{}^\alpha \;  ;\quad  {i\over 18} e^a
F_{a[3]} := F_{ab_1b_2b_3} \; ,  \quad  F^{[4]}:= F^{b_1b_2b_3b_4}
\; . \quad
\end{eqnarray}

The generalized covariant derivative is defined by the gravitino
supersymmetry transformations which may be written as
\begin{eqnarray}
\label{susyPsi}
\delta_{\varepsilon}\psi = {\cal
D}{\varepsilon}^\alpha = D {\varepsilon}^\alpha -
{\varepsilon}^\beta {t}{}_\beta{}^\alpha\, = d
{\varepsilon}^\alpha - {\varepsilon}^\beta {w}{}_\beta{}^\alpha\,=
d {\varepsilon}^\alpha - {\varepsilon}^\beta (\omega
+t){}_\beta{}^\alpha\, \;  \qquad
\end{eqnarray}
(up to quadratic fermion terms for type II, $D$=10). The Killing
spinor equation (\ref{Killing}) expresses the supersymmetry
invariance of a purely {\it bosonic} solution, since $\psi=0$
requires $\delta_{\varepsilon}\psi =0$. The preserved
supersymmetries are determined by the Grassmann odd functions
${\varepsilon}^\alpha(x)= \kappa^{{I}} \epsilon_{{I}}{}^\alpha(x)$
constructed from $k$ independent bosonic Killing spinors and the
fermionic parameters $\kappa^{{I}}$, ${{I}}=1, \ldots , k\,$.

The selfconsistency (integrability) condition for the Killing
spinor equation (\ref{Killing}), ${\cal D}{\cal D}
\epsilon_J{}^\alpha =0$, can be written
\cite{Duff+Kelly,FFP02,Duff03} in terms of the {\it generalized
curvature} ${\cal R} = dw - w\wedge w\;$ of the generalized
connection $w=\omega +t$ (eqs. (\ref{Killing}) and (\ref{t11D})
for $D$=11) as
\begin{eqnarray}\label{calR}
\epsilon_J{}^\beta  {\cal R}_\beta{}^{\alpha}  = 0  \quad (a)
\quad , & {\cal R}_\beta{}^{\alpha} = d\omega_\beta{}^{\alpha} -
\omega _\beta{}^{\gamma}\wedge \omega_\gamma{}^{\alpha} \;  \quad
(b) \quad .
\end{eqnarray}
The generalized curvature ${\cal R}_\beta{}^{\alpha}$ takes its
values in the Lie algebra ${\cal H}$ of the {\it generalized
holonomy group} $H$. This suggested to classify the partially
supersymmetric supergravity solutions by their generalized
holonomy \cite{Duff+Kelly,FFP02,Duff03}, which turned out useful
in the search for new solutions. It was found in
\cite{Hull03,P+T031} for D=11 and type II supergravity that $H$ is
a subgroup of $SL(32,\mathbb{R})$ (rather than
$GL(32,\mathbb{R})$).

Thus, a $\;k/32$ supersymmetric solution of the $D$=11 or of the
type II $D$=10 supergravity equations can be characterized by a
set of $k$ Killing spinors $\epsilon_{I}{}^\alpha$, ${I}=1, \ldots
, k$, obeying eq. (\ref{Killing}). Alternatively, we may use the
$\tilde{n}=32-k$ preonic spinors $\lambda_\alpha{}^r$, $r=1,
\ldots , \, \tilde{n}$, orthogonal to the Killing spinors, eq.
(\ref{epkXl=0}). Together, they make $\tilde{n}+k=32$ bosonic
spinors that can be used to define a {\it moving} $G$-{\it frame}
\cite{BPS03}.

By using the moving $G$-frame method it was found \cite{BPS03}
that the generalized curvature of a BPS preonic supergravity
solution should have the form
\begin{eqnarray}\label{cR31=}
{\cal R}_\beta{}^{\alpha}  =  dB^I \, \lambda_\beta \,
\epsilon_I{}^\alpha\; , \qquad {} \qquad \alpha= 1,\ldots , 32\; ,
\qquad
  I=1, \ldots , 31  \ , \quad
\end{eqnarray}
where $B^I= e^aB_a^I$ is a set of $31$ one-forms which determines
the non-trivial part of the $sl(32,\mathbb{R})$-valued generalized
connection $w_\beta{}^\alpha$ of the hypothetical preonic
solution, $w_\beta{}^{\alpha}  = \lambda_\beta \, B^I \,
\epsilon_I{}^\alpha - (dg\, g^{-1})_\beta{}^{\alpha}$. The
generalization of the $k=31$ equation (\ref{cR31=}) to the $k<31$
case \cite{BPS03}, also recovers the general statement
\cite{P+T03} that the generalized holonomy of a $k/32$
supersymmetric solution is the semidirect product
$SL(32-k,\mathbb{R}) \subset\!\!\!\!\!\!\times \,
\mathbb{R}^{k(32-k)}$, which gives $H=\mathbb{R}^{31}$ for the
preonic $k=31$ case.

Due to the presence of $\lambda_\beta$, eq. (\ref{cR31=}) solves
the selfconsistency conditions (\ref{calR}a) for the existence of
$31$ Killing spinors (\ref{Killing}). The appearance of the
Killing spinors $\epsilon_I{}^\alpha$ in the preonic generalized
curvature of eq. (\ref{cR31=}) follows from the fact
\cite{Hull03,P+T031} that the generalized holonomy group $H$ is a
subgroup of $SL(32, \mathbb{R})$ both for $D$=11 and type IIB
$D$=10 supergravities. This implies  ${\cal R}_\alpha{}^\alpha=0$
(zero trace), which is satisfied by (\ref{cR31=}) because of
(\ref{epkXl=0}). Furthermore, the actual form of the generalized
connection \cite{Hull03,P+T031} shows that they also are $sl(32,
\mathbb{R})$-valued, $w_\alpha{}^\alpha=0$, so that the
generalized structure group $G$ is also a subgroup of
$SL(32,\mathbb{R})$. This implies, after some algebra
\cite{BPS03}, that the preonic spinor of a preonic solution is
$G$-covariantly constant,
\begin{eqnarray}
\label{cDl31=0}
{\cal D}\lambda_{\beta} := D\lambda_{\beta}  + t_\beta{}^{\alpha}
\lambda_{\alpha}  := d\lambda_{\beta} + w_\beta{}^{\alpha}
\lambda_{\alpha} =0 \; , \qquad \alpha= 1,\ldots , 32\; , \qquad
  I=1, \ldots , 31  \, . \quad
\end{eqnarray}
Notice the difference in sign and position of indices between the
generalized covariant derivatives in eqs. (\ref{Killing}) and
(\ref{cDl31=0}). When the charge conjugation matrix
$C^{\alpha\beta}$ exists [IIA $D$=10, $D$=11], it may be used to
rise the spinorial indices but, since it is not `$G$-covariantly'
constant  (${\cal D}C^{\alpha\beta}:= - 2w^{[\alpha\beta]}\not=0$
unless the flux $F_{abcd}$ vanishes\footnote{It follows from
(\ref{t11D}) that $w_a^{[\alpha\beta]}\propto
F^{b_1b_2b_3b_4}(C\Gamma_{ab_1b_2b_3b_4})^{\alpha\beta}\;$, the
vanishing of which
 implies $F^{b_1b_2b_3b_4}=0$.}), the two `$G$-covariant'
 derivatives in (\ref{susyPsi}) and (\ref{cDl31=0})
 are different in general.

\section{No preonic solutions without $\alpha^\prime$ corrections in type II supergravities}

In $D=11$ the only fermionic field in the supergravity multiplet
is the gravitino $\psi= dx^\mu \psi_\mu(x)$, and the $k$ Killing
spinors describing supersymmetries of a $k/32$ bosonic solution of
the supergravity equations obey the differential equation
(\ref{Killing}) only. The $D$=10 type II supergravity theories
include, in addition, the $32$-component dilatino field
$\check{\chi}(x)$, which carries a reducible representation of
$Spin(1,9)$, and can be split into two Majorana-Weyl spinors with
different (IIA) or equal (IIB) chirality,
\begin{eqnarray}\label{IIADIL}
& \!\!\!\!\!\!\!\!\!\! \mathbf{IIA}  \;  \;
\check{\chi}^{\check{\alpha}}:= ({\chi}^{{\alpha}1}\, , \,
{\chi}^2_{{\alpha}})\; , \qquad \mathbf{IIB}  \; \;
\check{\chi}_{\check{\alpha}}:= ({\chi}^1_{{\alpha}}\; , \;
{\chi}^2_{{\alpha}}\,)\; , \qquad \check{\alpha}=1, \ldots 32\; ,
\; \alpha=1, \ldots 16  \; . \quad
\end{eqnarray}
The IIA and IIB dilatino supersymmetry transformation is algebraic,
$\delta_{susy} \check{\chi} = \check{\varepsilon} M$, where the
matrix $M$ is expressed, respectively, through the type IIA, IIB
supergravity fluxes.

Due to the presence of $\check{\chi}$, the Killing spinors
$\check{\epsilon}_I^{\check{\alpha}}$ associated with the preserved
supersymmetries of a type II purely bosonic solution
($\check{\psi}^{\check{\alpha}}_a=0$ and
$\check{\chi}^{\check{\alpha}}=0$) have to obey not only the
differential equation (\ref{Killing}), but also the {\it algebraic}
equation $\check{\epsilon}_I \, M\; =0$ that follows from
$\delta_{susy} \check{\chi}^{\check{\alpha}}=0$,
\begin{eqnarray}
\label{KEqGR+D}
 \textrm{type} \;\textrm {II}\, :\qquad  {\cal D}\check{\epsilon}_I := D
\check{\epsilon}_I - \check{\epsilon}_I {\check t}\, = 0\;  \quad
(a) \qquad & \check{\epsilon}_I \, M\; =0  \;    \quad (b)\qquad
(I=1, \ldots , k)\; .
\end{eqnarray}
Actually, the study of the algebraic Killing spinor equation
(\ref{KEqGR+D}b) led to establishing the absence of preonic
solutions, both in type IIB \cite{GGPR-IIB-06} and type IIA
supergravities \cite{BdAV06}. Applying the spinor moving $G$-frame
method of \cite{BPS03}, one finds that eq. (\ref{KEqGR+D}b) is
solved by
\begin{eqnarray}\label{M32=lxs}
\fbox{$\; M=\check{\lambda} \otimes \check{s}\;$}\;  ,  \quad
\textrm{namely \qquad
 IIA}\; : \; M_{\check{\beta}}{}^{\check{\alpha}}=
\check{\lambda}_{\check{\beta}} \; \check{s}^{\check{\alpha}} \; ,
\qquad \textrm{IIB}\; : \; M_{\check{\beta}\check{\alpha}}=
\check{\lambda}_{\check{\beta}} \; \check{s}_{\check{\alpha}} \quad,
\end{eqnarray}
 where $\check{\lambda}$ is the
$32$-component {\it preonic spinor} and $\check{s}$ is a generic
$32$-component spinor  which, at the final stage, has to be
expressed through the (IIA or IIB) fluxes (on-shell field
strengths) defining the $M$ matrix.
 For instance, for IIA supergravity
\begin{eqnarray}\label{IIAprl}
\hbox{IIA} \; : \qquad \check{\lambda}_{\check{\alpha}}:=
({\lambda}^1_{{\alpha}}\; , \; {\lambda}^{{\alpha}2})\; , \qquad
\check{s}^{\check{\alpha}}:= ({s}^{{\alpha}1}\; , \;
{s}^2_{{\alpha}})\; ,  \qquad \alpha=1,\ldots, 16\; , \qquad
\end{eqnarray}
and the $32\times 32$ matrix $M$ of eq. (\ref{KEqGR+D}b) reads
\begin{eqnarray}\label{M=IIA}
\hbox{IIA} \; :\quad  & M_{\check{\beta}}{}^{\check{\alpha}}=
{1\over 2}
\partial\!\!\!\!/\Phi \otimes \sigma_1 - {1\over 4} H\!\!\!\!/^{(3)} \otimes i\sigma_2
 + {3\over 8} e^{\Phi} R\!\!\!\!/^{(2)} \otimes \sigma_3
 + {1\over 8}R\!\!\!\!/^{(4)} \otimes \mathbb{I}_2 \; ,
\end{eqnarray}
where $\sigma_1 , \sigma_2 , \sigma_3$ are the standard $2\times
2$ Pauli matrices and the slashed variables, which involve the
$D=10$, $16\times 16$ Pauli matrices\footnote{In $D=10$,
$\sigma^a= \sigma^a_{\alpha\beta}$, ${\tilde \sigma}^a=
{\sigma^a}^{\alpha\beta}$, $a=0,1,\dots,9\,$. They obey
$\;\sigma^a{\tilde\sigma}^b+\sigma^b{\tilde\sigma}^a= 2
\eta^{ab}:= diag(+,-,\ldots , -)$ and
${\tilde\sigma}^a\sigma^b+{\tilde\sigma}^b \sigma^a= 2 \eta^{ab}$.
Then $\sigma^{ab...}:= (\sigma^{[a}\tilde{\sigma}^{b...]})$,
 $\tilde{\sigma}^{ab...}:= (\tilde{\sigma}^{[a}\sigma^{b...]})$,$\;$ {\it e.g.} $\sigma^{ab}:=
(\sigma^{[a}\tilde{\sigma}^{b]}){}_\alpha{}^\beta$  etc. } (also
denoted by the letter $\;\sigma$, $\;\sigma^a$) are given by
\begin{eqnarray}
\label{fl-NSNS} & \partial\!\!\!\!/\Phi := D_a\Phi
\sigma^a_{\alpha\beta}\; , \;\; H\!\!\!\!/^{(3)}= {1\over 3!}
H_{abc}\sigma^{abc} \; , \quad \tilde{\partial}\!\!\!\!/\Phi :=
D_a\Phi \tilde{\sigma}^{a\, \alpha\beta}\; , \;\;
\tilde{H}\!\!\!\!/^{(3)}= {1\over 3!}
H_{abc}\tilde{\sigma}^{abc}{}^{a\, \alpha\beta} \; , \quad
\\ \label{fl-RRIIA} & R\!\!\!\!/^{(2)}:= {1\over 2!}
R_{ab}(\sigma^{ab})_\alpha{}^\beta  =: -
\tilde{R}\!\!\!\!/^{(2)}{}^\beta{}_\alpha \; , \qquad
R\!\!\!\!/^{(4)}:= {1\over 4!}
R_{abcd}\sigma^{abcd}{}_\alpha{}^\beta =:
\tilde{R}\!\!\!\!/^{(4)}{}^\beta{}_\alpha\quad .
\end{eqnarray}
Thus, the IIA matrix M includes all the possible IIA fluxes,
namely,
\begin{eqnarray}
\label{fluxesIIA}
 R_2:=dC_1\; , \qquad R_4:=dC_3-C_1\wedge H_3\; , \qquad H_3:=dB_2 \; \qquad and
\qquad d\Phi \; .  \qquad
\end{eqnarray}

Eq. (\ref{M32=lxs}) restricts strongly both the fluxes in $M$ and
the preonic spinor $\check{\lambda}$. For IIB, it may be seen
\cite{BdAV06} that, when the quantum stringy corrections to the
supergravity equations and supersymmetry transformations are
ignored, the equation $M=\check{\lambda} \otimes \check{s}$
requires either $\check{\lambda}=0$ (no preonic spinor), which
implies a fully supersymmetric solution with $32$ rather than just
$31$ supersymmetries, {\it or} $\check{s}$=0 and, hence, $M=0$, in
which case all IIB fluxes but $R_5$ are zero. In this second case
($\lambda\neq 0$), it may be seen that the differential condition
(\ref{cDl31=0}) (or the Killing spinor equations (\ref{Killing})),
which involves the only non-zero flux $R_5$, implies that only an
{\it even} number of supersymmetries may be broken and thus no
preonic solutions, as previously found in \cite{GGPR-IIB-06}.

For type IIA the consequences of the algebraic equation
(\ref{KEqGR+D}b) are the same \cite{BdAV06}, but the condition
$M=0$ now implies that {\it all} fluxes are zero, in which case it
is known \cite{Duff03} that the only solutions with $k>16$
supersymmetries are the fully supersymmetric ones. Indeed, after
some algebra, one finds that eq. (\ref{M32=lxs}) implies that the
auxiliary spinor $\check{s}^{\check{\alpha}}=({s}^{{\alpha}1} \, ,
\,{s}^2_{{\alpha}})$ is related to the type IIA preonic spinor
($D$=10 Majorana spinor) $\check{\lambda}_{\check{\alpha}}:=
({\lambda}^1_{{\alpha}}\; , \; {\lambda}^{{\alpha}2})$ in
(\ref{IIAprl}) by  ${s}^{{\alpha}1} = a {\lambda}^{{\alpha}2}$,
${s}^2_{{\alpha}}= a {\lambda}^1_{{\alpha}}$ for some real
constant $a$, and that the the two Majorana-Weyl parts
${\lambda}^1_{{\alpha}}$ and ${\lambda}^{{\alpha}2}\;$ of
$\check{\lambda}_{\check{\alpha}}$ $\,$are restricted by a set of
equations including
\begin{eqnarray}
\label{dP=allV}
 \partial\!\!\!\!/\Phi:=\sigma^a_{{\alpha}\beta}D_a \Phi =
2a{\lambda}^1_{{\alpha}} {\lambda}^1_{\beta}\; ,
 \qquad
\tilde{\partial}\!\!\!\!/\Phi := \tilde{\sigma
}{}^{a{\alpha}\beta} D_a  \Phi = 2a{\lambda}^{{\alpha}2}
{\lambda}^{{\beta}2}\; . \qquad
\end{eqnarray}
These $16\times16$ matrix equations have only the trivial solution
$d\Phi=0$, $a{\lambda}^1_{{\alpha}} =0$,
$a{\lambda}^{{\alpha}2}=0$, and this implies the absence of
preonic supergravity solutions. To check that the solution of each
equation in (\ref{dP=allV}) is trivial, it is sufficient to notice
that the rank of the matrices in their {\it r.h.} sides is
obviously one or zero, while the the rank of those in their {\it
l.h} sides is 32 ($D_a\Phi$ timelike or spacelike), 16 (lightlike)
or zero. The only consistent choice is rank zero, which implies
the trivial solution. Thus, there is also a BPS preon conspiracy
in classical type IIA supergravity.

\section{Remarks on D=11, stringy corrections and preon conspiracy}

As far as the D=11 supergravity is concerned, a negative result on
the existence of simply connected preonic solutions was recently
reported in \cite{GGPR-11-06}. Eq. (\ref{cR31=}) \cite{BPS03},
expressing the selfconsistency condition for a preonic solution of
$D$=11 supergravity, was solved by the methods of spinorial
geometry (see refs. in \cite{GGPR-IIB-06,GGPR-11-06}) with the
result that such a 31/32 solution of the {\it free} supergravity
field equations actually has trivial generalized holonomy, ${\cal
R}_\beta{}^\alpha=0$, and hence is locally isomorphic to a 32/32
supersymmetric solution. Nevertheless, these results still allow
\cite{GGPR-11-06} for 31/32 solutions among not simply connected
manifolds\footnote{{\it Note added}. This possibility has just
been ruled out \cite{JFOF07}.}. Further, since the field equations
and the Bianchi identities were also used to reach the above
negative result, the presence of brane sources or of stringy
corrections to the supergravity equations might change the
conclusion.

To illustrate this let us notice \cite{BdAV06} that, in the light
of the results in \cite{M-thQC}, one may expect that the stringy
corrections to the supersymmetry transformations may modify the
algebraic structure of eqs. (\ref{dP=allV}). The essential point
is the appearance of some $Q^{\pm}_{abcde}$ contributions changing
(\ref{dP=allV}) to
\begin{eqnarray}
\label{dP=allVq?} &   \sigma^a_{{\alpha}\beta}D_a \Phi +
Q^-_{abcde} \sigma^{abcde}_{{\alpha}\beta}=
2a{\lambda}^1_{{\alpha}} {\lambda}^1_{\beta}  \; , \qquad
\tilde{\sigma}^{a\, {\alpha}\beta}D_a \Phi + Q^+_{abcde}
\tilde{\sigma}^{abcde\;{\alpha}\beta} =  2a{\lambda}^{2{\alpha}}
{\lambda}^{2\beta} \; .
 \qquad
\end{eqnarray}
With a general $Q^{\pm}_{abcde}$ tensor, the algebraic structure
of the {\it l.h.} sides of these equations correspond to the
decompositions of general symmetric $16\times 16$ matrices into
irreducible $SO(1,9)$ tensors. Thus, there are no restrictions on
their rank and, hence, these equations may have nontrivial
solutions.
\medskip

    The BPS preon conspiracy, {\it i.e.} the absence of
$31/32$ supersymmetric solutions for the $D$=11 and $D$=10 type II
classical supergravities \cite{GGPR-IIB-06,BdA06-J,GGPR-11-06}
discussed here, is consistent with the preonic conjecture (one
might think, by way of an analogy, of quark confinement, which
does not preclude the existence of quarks). Introduced as
M-theoretical objects \cite{BPS01}, the `observation' of BPS
preons might require considering M-theoretic/stringy corrections
to the supergravity equations; they could appear already at the
next $\propto \alpha^\prime{}^3$ approximation. This would imply
that BPS preons are intrinsically quantum objects which cannot be
`seen' in a classical supergravity description.

 The present challenge for the preonic hypothesis is to exhibit
its real usefulness in the study of String/M-theory. At present,
it provides an algebraic classification of all BPS states, and the
preon-inspired moving $G$-frame method has been useful for
studying supergravity solutions. To conclude, it is also
interesting to note that dynamical models for the $D$=4,6,10
counterparts of (pointlike) BPS preons ($n$=4, 8 and 16, see below
(\ref{M-alg})) describe massless conformal higher spin theories
(see \cite{30/32, BBdAST04, DimaHS, BdA06-J} and refs. therein).

\bigskip
{\bf Acknowledgements}. Partial support from the Spanish Ministry
of Education and Science (FIS2005-02761) and EU FEDER funds, the
Generalitat Valenciana, the Ukrainian State Fund for Fundamental
Research (383), the INTAS project 2006-7928 and the EU `Forces
Universe' network MRTN-CT-2004-005104, is gratefully acknowledged.


\begin{thebibliography}{[99]}

\bibitem{BPS01}
I.\,A.~Bandos, J.\,A. de Azc\'arraga, J.\,M.~Izquierdo and
J.\,Lukierski, {\it BPS states in M-theory and twistorial
constituents},
  Phys.\ Rev.\ Lett.\  \textbf{86}, 4451-4454 (2001)
  [hep-th/0101113].

\bibitem{SGsol}
 M.~J.~Duff, B.~E.~W.~Nilsson and C.~N.~Pope,
 {\it Kaluza-Klein supergravity},
  Phys.\ Rept.\  {\bf 130} (1986) 1-142;
M.~J.~Duff, R.~R.~Khuri and J.~X.~Lu,
 {\it String solitons},
  Phys.\ Rept.\  {\bf 259}, 213-326 (1995)
  [hep-th/9412184];
 K.~S.~Stelle,
 {\it BPS branes in supergravity},
in {\it High energy physics and cosmology}, Trieste 1997, {\it The
ICTP Series in Theoretical Physics} {\bf 14}, E. Gava, et al.
eds., A. Masiero, K.S. Narain, S. Randjbar-Daemi, G. Senjanovic,
A. Smirnov and Q Shafi Eds., World Scientific, (1998), pp. 29-127
[hep-th/9803116].

\bibitem{JK-G83-Hull84}
J.~Kowalski-Glikman, {\it Vacuum States In Supersymmetric
Kaluza-Klein Theory},
  Phys.\ Lett.\ B {\bf 134}, 194-196 (1984);
  C.~M.~Hull,
  {\it Exact pp wave solutions of 11-dimensional supergravity},
  Phys.\ Lett.\ B {\bf 139}, 39-41 (1984).

\bibitem{BL98}
I. Bandos and J. Lukierski, {\it Tensorial central charges and new
superparticle models with fundamental spinor coordinates}, Mod.
Phys. Lett. {\bf 14}, 1257-1272 (1999) [hep-th/9811022];
{\it New superparticle models outside the HLS supersymmetry
scheme},
  Lect.\ Notes Phys.\  {\bf 539}, 195 (2000) [hep-th/9812074].

\bibitem{G+H99}
J.P. Gauntlett and C.M. Hull, {\it BPS States with Extra
Supersymmetry}, JHEP {\bf 0001}, 004 (2000) [hep-th/9909098].

\bibitem{BPS03}
 I.~A.~Bandos, J.~A.~de Azc\'{a}rraga, J.~M.~Izquierdo, M.~Pic\'on and O.~Varela,
  {\it On BPS preons, generalized holonomies and D = 11 supergravities},
  Phys.\ Rev.\  {\bf D69}, 105010 (2004)
  [hep-th/0312266].

\bibitem{ppW}
M.~Cveti\v c, H.~L\"u and C.N.~Pope, {\it M-theory PP-waves,
Penrose limits and supernumerary supersymmetries},
Nucl.~Phys.~{\bf B644}, 65 (2002) [hep-th/0203229]; \,

J.P.~Gauntlett and C.M.~Hull, {\it PP-waves in 11-dimensions with
extra supersymmetry}, JHEP 0206, 013 (2002) [hep-th/0203255]; \,

J.~Michelson, {\it (Twisted) toroidal compactification of
pp-waves}, Phys.\ Rev. {\bf D66}, 066002 (2002) [hep-th/0203140];
{\it A pp-Wave With 26 Supercharges}, Class.~Quant.~Grav. {\bf
19}, 5935 (2002) [hep-th/0206204]; \,

I.~Bena and R.~Roiban, {\it A supergravity pp-wave solutions with
28 and 24 supercharges}, Phys.~Rev.~{\bf D67}, 125014 (2003)
[hep-th/0206195].

\bibitem{Duff02}
M.~J.~Duff, {\it M-theory on manifolds of G(2) holonomy: The first
twenty years}, hep-th/0201062.

\bibitem{GGPR-IIB-06}
  U.~Gran, J.~Gutowski, G.~Papadopoulos and D.~Roest,
  {\it N = 31 is not IIB}, hep-th/0606049.

\bibitem{BdAV06}
  I.\,A.~Bandos, J.\,A.~de Azc\'arraga and O.~Varela,
{\it On the absence of BPS preonic solutions in IIA and IIB
supergravities}, JHEP {\bf 0609}, 009 (2006)
  [hep-th/0607060].

\bibitem{GGPR-11-06}
  U.~Gran, J.~Gutowski, G.~Papadopoulos and D.~Roest,
{\it N = 31, D = 11}, hep-th/0610331.

\bibitem{30/32}
 I.~A.~Bandos, J.~A.~de Azc\'arraga, M.~Pic\'on and O.~Varela,
{\it D = 11 superstring model with 30 kappa-symmetries and 30/32
BPS states  in an extended superspace},
  Phys. Rev. {\bf D69}, 085007 (2004) [hep-th/0307106];
 I.A.~Bandos,
{\it BPS preons and tensionless super-p-branes in generalized
superspace}, Phys. Lett. {\bf B558}, 197-204 (2003)
[hep-th/0208110];
{\it BPS preons in supergravity and higher spin theories: An
overview from  the hill of twistor appraoch},
  AIP Conf.\ Proc.\ {\bf 767}, 141-171 (2005)
  [hep-th/0501115].

\bibitem{BdA06-J}
I.~A.~Bandos and J.A. de Azc\'arraga,
   {\it BPS preons and higher spin theory in D=4,6,10}, to appear in
{\it Quantum, Super and Twistors}, Proc. of the XXII Max Born
Symposium, 27-29 Sep. 2006, Wroclaw (Poland), hep-th/0612277.

\bibitem{Duff+Kelly}  M.~J.~Duff and K.~S.~Stelle,
{\it Multi-membrane solutions of D = 11 supergravity},''
  Phys.\ Lett.\ {\bf B253}, 113 (1991).

\bibitem{FFP02}
J.~Figueroa O'Farrill and G.~Papadopoulos, {\it Maximally
supersymmetric solutions of ten and eleven-dimensional
supergravities}, JHEP  {\bf 0303}, 048 (2003) [hep-th/0211089]

\bibitem{Duff03}
M.~J.~Duff and J.~T.~Liu, {\it Hidden spacetime symmetries and
generalized holonomy in M-theory}, Nucl.\ Phys.\ B {\bf 674},
217-230 (2003) [hep-th/0303140].

\bibitem{Hull03}
C.~Hull, {\it Holonomy and Symmetry in M-theory}, hep-th/0305039.

\bibitem{P+T031}
G.~Papadopoulos and D.~Tsimpis, {\it The holonomy of IIB
supercovariant connection}, Class.\ Quant.\ Grav.\  {\bf 20},
L253-L258 (2003)  [hep-th/0307127].

\bibitem{GP02}
J.~P.~Gauntlett and S.~Pakis, {\it The geometry of $D=11$ Killing
spinors}, JHEP {\bf 0304}, 039 (2003) [hep-th/0212008].

\bibitem{P+T03}
G.~Papadopoulos and D.~Tsimpis, {\it The holonomy of the
supercovariant connection and Killing spinors}, JHEP {\bf 0307},
018 (2003) [hep-th/0306117].

\bibitem{M-thQC}
H. Lu, C.N. Pope, K.S. Stelle and P.K. Townsend, {\it String and
M-theory deformations of manifolds with special holonomy}, JHEP
{\bf 0507}, 075 (2005) [hep-th/0410176];
\\
H.~Lu, C.~N.~Pope and K.~S.~Stelle, {\it Generalised holonomy for
higher-order corrections to supersymmetric backgrounds in string
and M-theory}, Nucl. Phys. {\bf B741}, 17-33 (2006)
[hep-th/0509057].

\bibitem{BBdAST04}
I.A.~Bandos, X.~Bekaert, J.~A.~de Azc\'arraga, D. Sorokin and M.
Tsulaia, {\it Dynamics of higher spin fields and tensorial space},
JHEP {\bf 0505}, 031 (2005) [hep-th/0501113];
I.~Bandos, P.~Pasti, D.~Sorokin and M.~Tonin, {\it Superfield
theories in tensorial superspaces and the dynamics of higher spin
fields},  JHEP {\bf 0411}, 023 (2004) [hep-th/0407180].

\bibitem{DimaHS}
 D.~Sorokin, {\it Introduction to the classical theory of higher spins},
  AIP Conf.\ Proc.\  {\bf 767}, 172-202 (2005)  [hep-th/0405069].

\bibitem{JFOF07}
 J.~Figueroa-O'Farrill and S.~Gadhia,
  {\it M-theory preons cannot arise by quotients}, hep-th/0702055.

\end{thebibliography}
\end{document}